\documentstyle[epsf,preprint,aps]{revtex}
\draft \preprint{SNUTP 02/024}
\begin{document}
\title{\Large\bf More self-tuning solutions
with $H_{MNPQ}$ }
\author{Jihn E. Kim and Hyun Min Lee}
\address{ Department of Physics and Center for Theoretical
Physics, Seoul National University, Seoul 151-747, Korea}
\maketitle

\begin{abstract}
We find more self-tuning solutions by introducing a general form
for Lagrangian of a 3-index antisymmetric tensor field $A_{MNP}$
in the RS II model. In particular, for the logarithmic Lagrangian,
$\propto\log(-H^2)$, we obtained a closed form weak self-tuning
solution.
\end{abstract}

\newpage
\section{Introduction}

Recently, the self-tuning solutions have been attempted toward
solutions of the cosmological constant problem. The self-tuning
solutions can be broadly classified to

$(i)$ {\it weak self-tuning solutions:} this class requires just the existence
of the flat space solution in 4 dimensional(4D) space-time, and

$(ii)$ {\it strong self-tuning solutions:} this class allows only the 4D
flat solution without the possibility of 4D curved space solutions.

Of course, a strong self-tuning solution can be considered as a solution of
the cosmological constant problem. However, the recent attempts toward a
strong self-tuning solution by Kachru {\it al.}
\cite{kachru} has not been successful due to a nonlocalizable
gravity or resurrection of the fine-tuning problem after curing the
singularity\cite{nilles}. It seems that a strong self-tuning solution is
difficult to realize at present.

On the other hand, the weak self-tuning solutions are easier to
realize. Because the weak self-tuning solutions do not forbid de
Sitter or anti de Sitter space solutions, it is necessary to
supply an additional principle to choose a flat one out of
numerous possibilities. Witten argued that probably the boundary
of different phases is chosen\cite{witten}, and Hawking argued the
Euclidian quantum gravity gives the maximum probability for the
flat universe\cite{hawking}. We note that the weak self-tuning
solution is a big progress. In this regard, note that in 4D a
nonvanishing cosmological constant never allows a flat solution,
and one has to fine-tune the 4D cosmological constant at zero to
have a flat space solution. However, if a weak self-tuning
solution is present, then it is a matter of choosing the flat
space solution out of numerous possibilities. Witten and Hawking
used the 4-form field strenth\cite{townsend}
$H_{\mu\nu\rho\sigma}$ to show the weak self-tuning solution.
However, the 4-form field in 4D is not a dynamical field and a
weak self-tuning solution cannot be realized in an evolving
universe. Even if a flat solution is chosen in the early universe,
phase transitions at later epoches can add vacuum energies and can
transform  the flat solution to curved ones. If the weak
self-tuning solution is realized with a dynamical field with an
undetermined integration constant(UIC) $c$ which determines the
profile of the dynamical field, then addition of vacuum energy can
change the profile of the field so that the resulting space-time
remains flat. Therefore, it is necessary to have the weak
self-tuning solution with a dynamical field. If a dynamical field
plays the required role, it is better for it to be a massless
scalar so that it affects the whole space-time. Indeed, a weak
self-tuning solution having these properties was found
\cite{kkl1,kkl2} in a 5D Randall-Sundrum II type model\cite{rs2},
using a 5D 4-form field strength $H_{MNPQ}$. In 5D, the 4-form
field strength has one massless scalar. The Lagrangian considered
is $1/H^2$ where $H^2 =H_{MNPQ}H^{MNPQ}$. At low energy, this
Lagrangian can be considered as an effective one. The solution
found in \cite{kkl1} can be considered at least as the existence
proof for the weak-self-tuning solution.

The 4-form field strength $H_{MNPQ}$, having one dynamical field
in 5D and being massless due to the gauge symmetry, might be a key
ingredient to the self-tuning solutions. In this paper, therefore,
we consider more general functions of $H^2$ allowing weak
self-tuning solutions.

\section{General Form for Lagrangian with $H^2$ in 5D}

Introducing a 3-index antisymmetric tensor field $A_{MNP}$ in the RS
II model\cite{rs2},
let us consider a general form of Lagrangian with its field strength
$H_{MNPQ}=\partial_{[M}A_{NPQ]}$. Then the 5D action reads
\begin{eqnarray}
S=\int d^4 x dy\sqrt{-g}\bigg(\frac{1}{2}R-\Lambda_b+K(H^2)\bigg)
+\int_{y=0} d^4 x \sqrt{-g_4}(-\Lambda_1)
\end{eqnarray}
where $H^2=H_{MNPQ}H^{MNPQ}$, $g$ and $g_4$ are 5D and 4D metric
determinants, and $\Lambda_b$ and $\Lambda_1$ are bulk and brane
cosmological constants. Here we notice that a surface term is
needed for the well-defined variation of the action with respect
to $A_{MNP}$,
\begin{eqnarray}
S_{surf}=-2\int d^4 x dy\partial_M\bigg(\sqrt{-g}
\frac{\partial K(H^2)}{\partial H^2}H^{MNPQ}A_{NPQ}\bigg).
\end{eqnarray}

Then, the energy-momentum tensor becomes
\begin{eqnarray}
T_{MN}=-\Lambda_b g_{MN}-\frac{\sqrt{-g_4}}{\sqrt{-g}}g_{\mu\nu}
\delta_M^\mu\delta_N^\nu\Lambda_1\delta(y)+T^H_{MN}
\end{eqnarray}
where the contribution coming from $A_{MNP}$ is given by
\begin{eqnarray}
T^H_{MN}=K(H^2)g_{MN}-8\frac{\partial K(H^2)}{\partial H^2}H_{MPQR}H_N\,^{PQR}.
\end{eqnarray}

To obtain a 4D flat solution, let us take the ansatze for the metric and the
field strength as
\begin{eqnarray}
ds^2=\beta^2(y)\eta_{\mu\nu}dx^\mu dx^\nu+dy^2, \\
H_{\mu\nu\rho\sigma}=\sqrt{-g}\epsilon_{\mu\nu\rho\sigma}f(y), \ \ \ \ \
H_{5\mu\nu\rho}=0.
\end{eqnarray}
Then, the relevant Einstein equations and the field equation are
\begin{eqnarray}
3\bigg(\frac{\beta'}{\beta}\bigg)^2
+3\bigg(\frac{\beta^{\prime\prime}}{\beta}\bigg)&=&-\Lambda_b
-\Lambda_1\delta(y)+K(H^2)+8\cdot 3!\frac{\partial K(H^2)}{\partial H^2} f^2,
\label{mueq}\\
6\bigg(\frac{\beta'}{\beta}\bigg)^2
&=&-\Lambda_b+K(H^2),\label{master}
\end{eqnarray}
and
\begin{eqnarray}
\partial_M\bigg(\sqrt{-g}
\frac{\partial K(H^2)}{\partial H^2}H^{MNPQ}\bigg)=0\label{field}
\end{eqnarray}
where the argument in $K^2(H^2)$ is $H^2=-4!f^2$.  The bulk
equations of motion given above, which gives the condition that
Eq.~(\ref{master}) reproduces Eq.~(\ref{mueq}) in the bulk or the
Bianchi identity(dH=0), requires $f(y)\propto \beta^{-4}$.

\section{Self-Tuning Solution with $\ln (-H^2)$}

We can rewrite Eq.~(\ref{master}) as
\begin{eqnarray}
|\beta'|=\sqrt{-\frac{\Lambda_b}{6}\beta^2+\frac{\beta^2}{6} K(H^2)},
\label{restrict}
\end{eqnarray}
with $H^2=-2\cdot 4! Q\beta^{-8}$ where $Q$ is a positive
integration constant of Eq.~(\ref{field}). In dS and AdS spaces
with the curvature $\lambda$ ($>0$ for dS and $<0$ for AdS),
$\lambda$ is added in the square bracket of (\ref{restrict}). For
the existence of the self-tuning solution, a necessary condition
is that $\beta'$ should go to zero as $\beta$ goes to
zero\cite{kkl2},
\begin{equation}\label{flatcondition}
{\rm Flat\ condition}:\ \ \beta^\prime\rightarrow 0\ \mbox{ as }
\beta\rightarrow 0,
\end{equation}
which restricts the form of the functional $K(H^2)$. Note that the
de Sitter space solution gives a nonzero $\beta^\prime$ as
$\beta\rightarrow 0$, and anti de Sitter space solution gives a
nonzero $\beta$ as $\beta^\prime\rightarrow 0$. Indeed, if a
solution satisfying this condition is found and it also allows a
localizable gravity, it is an acceptable weak self-tuning
solution. Therefore, up to an acceptable 4D gravity, this
condition becomes the necessary and sufficient condition for the
existence of a 4D flat solution. For instance, for a single term
with power form of $H^2$ it was shown that only a negative power
of $H^2$ satisfies such a necessary and sufficient
condition\cite{kkl1,kkl2}. There can exist more solutions
satisfying this condition. In this paper, we are interested in
infinite series of $H^2$ which sum up to interesting elementary
functions and satisfy the above condition (\ref{flatcondition}).
In this section, we show another closed form weak self-tuning
solution with the logarithmic function. In the dual picture, the
bulk Lagrangian can be again a logarithmic function.

\subsection{Logarithmic function}

Let us take $K(H^2)$ as a logarithmic function of $H^2$ as
\begin{eqnarray}
K(H^2)=V{\rm log}\bigg(-\frac{H^2}{2\cdot 4!}\bigg)
\end{eqnarray}
where $V>0$ is needed for producing a conventional kinetic term for $A_{MNP}$
as a perturbation around the background solution.
Then, it is easy to check that this function satisfies the needed condition
because $\beta^2 K(H^2)=V\beta^2 (-8{\rm log}\beta+\cdots)$ vanishes
as $\beta\rightarrow 0$.

With this logarithmic function, the Einstein equations become
\begin{eqnarray}
3\bigg(\frac{\beta'}{\beta}\bigg)^2
+3\bigg(\frac{\beta^{\prime\prime}}{\beta}\bigg)&=&-\Lambda_b
-\Lambda_1\delta(y)+V({\rm log}Q-8{\rm log}\beta-2), \\
6\bigg(\frac{\beta'}{\beta}\bigg)^2 &=&-\Lambda_b+V({\rm
log}Q-8{\rm log}\beta).
\end{eqnarray}
Then, we obtain a 4D flat solution consistent with $Z_2$ symmetry as
\begin{eqnarray}
\beta(y)={\rm exp}\bigg[-\frac{\Lambda_b}{8V}+\frac18 \log Q
-\frac{V}{3}(|y|+c)^2\bigg] \label{flatsoln}
\end{eqnarray}
where the argument in the exponent contains a combination
\begin{eqnarray}
\kappa=-\frac{\Lambda_b}{6}+\frac{V}{6}{\log}Q.
\end{eqnarray}
The undetermined integration constant(UIC) $c$ is determined by
the boundary condition at the brane
\begin{eqnarray}
\frac{\beta'}{\beta}\bigg|_{y=0^+}=-\frac{\Lambda_1}{6},\label{bc}
\end{eqnarray}
as
\begin{eqnarray}
c=\frac{\Lambda_1}{4V}.
\end{eqnarray}

Therefore, we find that there exist regular flat solutions for an
arbitrary set of ($\Lambda_1, V$), irrespective of the value of
$\kappa$. Moreover, the 4D Planck mass becomes finite even for the
non-compact extra dimension,
\begin{eqnarray}
M^2_P&=&M^3\int^{\infty}_{-\infty}dy\, \beta^2(y)
=M^3 e^{2\kappa/v}\int^{\infty}_{-\infty}dy\, e^{-\frac{v}{2}(|y|+c)^2}
\nonumber \\
&=&2M^3 e^{2\kappa/v}\sqrt{\frac{\pi}{2v}}
\bigg(1-{\rm erf}\sqrt{\frac{v}{2}}c\bigg)
\end{eqnarray}
where $v=\frac43 V$, and erf denotes the error function
\begin{eqnarray}
{\rm erf}\,x=\frac{2}{\sqrt{\pi}}\int^x_0 dy\, e^{-y^2}.
\end{eqnarray}

\subsection{Dual Description }

Let us consider the above Lagrangian in the dual picture. In the
dual picture, a brane coupling of the dual scalar field can be
introduced easily.

In the dual description of a form field, the equation of motion
for a form field is transformed into the Bianchi identity for its
dual form field and the Bianchi identity of the form field becomes
the equation of motion of the dual form field. Thus, the equation
of motion for $A_{MNP}$, Eq.~(\ref{field}), becomes the Bianchi
identity for the dual scalar field $\sigma$,
\begin{eqnarray}
{\rm d}\sigma=\ *{\rm H}\cdot \frac{\partial K(H^2)}{\partial
H^2}.
\end{eqnarray}
That is, with a dual transformation such as\footnote{The dual
description of $1/H^2$ was considered in~\cite{kschoi}.}
\begin{eqnarray}
H_{MNPQ}=-\frac{1}{4!}\sqrt{-g}\epsilon_{MNPQR}
\frac{\partial^R\sigma}{(\partial\sigma)^2},
\end{eqnarray}
the 5D action including a brane coupling becomes
\begin{eqnarray}
S=\int d^4 x dy\sqrt{-g}\bigg(\frac{1}{2}R-\Lambda_b
-V[2+\log(2\cdot 4!(\partial\sigma)^2)]\bigg)
+\int_{y=0} d^4 x\sqrt{-g_4}(-\Lambda_1 U(\sigma)).
\end{eqnarray}
Therefore, the equation of motion for $\sigma$ becomes
\begin{eqnarray}
2V\partial_M\bigg(\sqrt{-g}\frac{\partial^M\sigma}{(\partial\sigma)^2}\bigg)
=\sqrt{-g_4}\Lambda_1\frac{dU}{d\sigma}\delta(y)
\end{eqnarray}
where a non-constant brane coupling $U(\sigma)$ implies that the
Bianchi identity for $A_{MNP}$ is not satisfied any more at the
brane. On the other hand, the energy-momentum tensor coming from
the dual field becomes
\begin{eqnarray}
T^\sigma_{MN}=V\bigg(-[2+\log(2\cdot 4!(\partial\sigma)^2)] g_{MN}
+2\frac{\partial_M\partial_N\sigma}{(\partial\sigma)^2}\bigg).
\end{eqnarray}

Then, assuming 4D Poincar$\acute{\rm e}$ invariance as
\begin{eqnarray}
ds^2=\beta^2(y)\eta_{\mu\nu}dx^\mu dx^\nu+dy^2, \ \ \ \sigma=\sigma(y),
\end{eqnarray}
we obtain the Einstein's equations and the field equation as
\begin{eqnarray}
3\bigg(\frac{\beta'}{\beta}\bigg)^2
+3\bigg(\frac{\beta^{\prime\prime}}{\beta}\bigg)&=&-\Lambda_b
-\Lambda_1U(\sigma)\delta(y)-V(2+\log(2\cdot 4!\sigma^{\prime 2})) \\
6\bigg(\frac{\beta'}{\beta}\bigg)^2
&=&-\Lambda_b-V\log(2\cdot 4!\sigma^{\prime 2})
\end{eqnarray}
and
\begin{eqnarray}
2V \bigg(\frac{\beta^4}{\sigma'}\bigg)'=\beta^4\Lambda_1
\frac{dU}{d\sigma}\delta(y).
\end{eqnarray}

Consequently, with $2\cdot 4!\sigma^{\prime 2}=\beta^8/Q$, the
bulk solution for the metric is the same as Eq.~(\ref{flatsoln})
in the case without a scalar coupling. But, due to the presence of
the scalar coupling, we get different boundary conditions for the
metric and the dual field at the brane:
\begin{eqnarray}
\frac{\beta'}{\beta}\bigg|_{y=0^+}&=&-\frac{\Lambda_1}{6}U(\sigma(0)), \\
\frac{1}{\sigma'}\bigg|_{y=0^+}&=&\frac{\Lambda_1}{4V}\frac{dU}{d\sigma}
(\sigma(0)).
\end{eqnarray}
Therefore, we need two consistency conditions, arising from the
existence of the brane and the scalar coupling,
\begin{eqnarray}
c&=&\frac{\Lambda_1}{4V}\,U(\sigma(0)),\label{BC1}\\
\pm\gamma
e^{[\frac{4V}{3}c^2+\Lambda_b/(2V)]}&=&\frac{\Lambda_1}{4V}\,
\frac{dU}{d\sigma}(\sigma(0))\label{BC2}
\end{eqnarray}
where $\gamma\equiv \sqrt{2\cdot 4!}$. That is, the condition for the scalar
coupling at the brane becomes
\begin{eqnarray}\label{lambdaphi}
\frac{1}{U}\frac{dU}{d\sigma}(\sigma(0))=\pm \frac{\gamma}{c}
e^{[\frac{4V}{3}c^2+\Lambda_b/(2V)]},
\end{eqnarray}
plus one of (\ref{BC1}) and (\ref{BC2}).

\subsection{Curved space solutions}

The de Sitter and anti de Sitter space solutions are parametrized
by the curvature $\lambda$ ($\lambda>0$ for dS and $\lambda<0$ for
AdS). The relevant equations in the dual picture are
\begin{eqnarray}
&3\left(\frac{\beta^{\prime\prime}}{\beta}\right)+\left(\frac{\beta^\prime}{
\beta}\right)^2-3\lambda\beta^{-2}=-\Lambda_b-U\Lambda_1\delta(y)
+V(-8\ln\beta+\ln\tilde Q-2) \\
&6\left(\frac{\beta^\prime}{\beta}\right)^2-6\lambda\beta^{-2}=-\Lambda_b
+V(-8\ln\beta+\ln\tilde Q)\label{yy}
\end{eqnarray}
where we used $\tilde Q$ to represent the charge. In terms of
$A(y)=\ln \beta(y)$, Eq.~(\ref{yy}) becomes
\begin{equation}
\frac{dA}{\sqrt{-\frac{\Lambda_b}{6}+\lambda e^{-2A}-\frac43
VA+\frac16 V\ln\tilde Q}}=dy,
\end{equation}
which can be integrated to give, for the case with the $Z_2$
symmetry,
\begin{equation}\label{dsform}
||y|+{\rm constant}|=\int^A
\frac{dA}{\sqrt{-\frac{\Lambda_b}{6}+\lambda e^{-2A}-\frac43
VA+\frac16 V\ln\tilde Q}}.
\end{equation}
The left-hand side of Eq.~(\ref{dsform}) contains an integration
constant and $\lambda$, which can be formally expressed as
$F(|y|;{\lambda}(\sigma_0),\cdots)$ where
$\sigma_0=\sigma(0)$.\footnote{ For $\lambda=0$, integral of
Eq.~(\ref{dsform}) gives $(3/2V)\sqrt{\kappa^2-A}$ or we obtain
$A=\frac34 \frac{\kappa^2}{V}-\frac13 V(|y|+c)^2$ which is
identical to Eq.~(\ref{flatsoln}).} If the scalar-brane coupling
is assumed, for illustration, as $U(\sigma)=e^{b\sigma}$ with a
parameter $b$, we obtain a curved space boundary condition,
similar to (\ref{lambdaphi}) of the flat case, as
\begin{equation}\label{dsBC2}
\frac{1}{\gamma} \exp\left(\frac{3\tilde\kappa^2}{V}-\frac43
V\tilde c^2\right)\cdot \sqrt{\frac{\tilde c^2}{\tilde
Q}+\frac{9\lambda}{4V^2\tilde Q} e^{-(3\tilde
\kappa^2/2V)+(2V\tilde c^2/3)}}=\frac{1}{b}
\end{equation}
where tilde denote the curved space constants. There is another
condition similar to (\ref{BC1}),
\begin{equation}\label{dsBC1}
\sqrt{\tilde\kappa^2+\lambda e^{-2A(0)}-\frac43
VA(0)}=\frac{\Lambda_1}{6}e^{b\sigma(0)}.
\end{equation}
For $\lambda=0$, the relation (\ref{dsBC2}) reduces to the flat
case $(c/\gamma\sqrt{Q})\exp[(3\kappa^2/V)-\frac43 Vc^2]=1/b$.  We
treated $b$ as a parameter. The constant $\tilde Q$ is determined
by the finite 4D Newton constant. $\tilde c$ is an integration
constant. There are two boundary conditions (\ref{dsBC2}) and
(\ref{dsBC1}), but the constants to be determined are three:
$\tilde c, \lambda$ and $\sigma(0)$. Note that $\sigma(0)$ is an
additional constant. Anyway, there exist curved space solutions in
our case, with a parameter undetermined. It is different from
Kachru {\it et al.} case where their solution imposes a specific
value for $b$ such that two equations are consistent only for
$\lambda=0$. To determine $\lambda$ uniquely in our case, we need
another condition.

\section{Other Self-Tuning Solutions}

It can be shown that there also exist self-tuning solutions for
the exponential functions and their some linear combinations such
as
\begin{eqnarray}
(1)~~:~~ K(H^2)&=&V {\rm exp}\bigg(q\frac{2\cdot 4!}{H^2}\bigg),\ \ Vq>0 ,\\
(2)~~:~~ K(H^2)&=&-V {\rm exp}\bigg(p\frac{H^2}{2\cdot 4!}\bigg),\
\ V>0 \mbox{ and } p>0,\\
(3)~~:~~ K(H^2) &=& -V\tanh \left(r\frac{H^2}{2\cdot 4!
}\right),\ \ Vr>0,\\
(4)~~:~~ K(H^2) &=& V\coth\left( s\frac{H^2}{2\cdot 4!}\right),\ \
Vs>0.
\end{eqnarray}
For the 4D flat solution, there exists a limit that $\beta^2
K(H^2)$ goes to zero for $\beta\rightarrow 0$ in both cases, which
is consistent with Eq.~(\ref{restrict}). When we expand the case
(1) in power series of $1/H^2$, we can obtain a class of general
self-tuning solutions previously argued with negative powers of
$H^2$ in Ref.~\cite{kkl1}. On the other hand, for cases (2) to
(4), we find it interesting that there exists a self-tuning
solution for the infinite sum of positive powers of $H^2$, but we
know that a finite sum does not allow a solution\cite{kkl2,kkl3}.

Then, to obtain additional self-tuning solutions, with the
boundary condition Eq.~(\ref{bc}), we only need to solve one
equation, for example, for cases (1) and (2), as follows
\begin{eqnarray}
(1)~~:~~6\bigg(\frac{\beta'}{\beta}\bigg)^2=-\Lambda_b+V e^{-q\beta^8/Q},\\
(2)~~:~~6\bigg(\frac{\beta'}{\beta}\bigg)^2=-\Lambda_b-V
e^{-pQ\beta^{-8}}.
\end{eqnarray}
For the case (2), we find that there exists a self-tuning solution
only for $\Lambda_b<0$.

The self-tuning functionals $K(H^2)$ considered in the previous
section and the current section can be generalized to those with
their argument replaced by some polynomial of $H^2$
\begin{eqnarray}
K(H^2)&=&V{\rm log}\bigg[\sum^{N_2}_{n=-N_1}
a_n\bigg(-\frac{H^2}{2\cdot 4!}\bigg)^n\bigg] \\
&=& V\bigg[{\rm log}\bigg(\sum^{N_2}_{n=-N_1} a_n Q
^n\beta^{8(N_2-n)}\bigg)-8N_2 {\rm log}\beta\bigg]
\end{eqnarray}
where $a_n$ are arbitrary constant coefficients
and $N_{1,2}$ are assumed to be arbitrary natural numbers.
Likewise, the case with the exponential form can be also generalized to
\begin{eqnarray}
K(H^2)&=&V {\rm exp}\bigg[\sum^{N_2}_{n=-N_1}a_n
\bigg(-\frac{H^2}{2\cdot 4!}\bigg)^n\bigg] \\
&=&V {\rm exp}\bigg[\sum^{N_2}_{n=-N_1}a_n Q^n \beta^{-8n}\bigg]
\end{eqnarray}
where all $a_n$ with $n>0$ should be negative for $\beta^2 K(H^2)$
to be zero as $\beta$ vanishes.

\section{Conclusion}

We obtained more weak self-tuning solutions of the cosmological
constant in RS-II type models with $H_{MNPQ}$. For many cases of
functions of $H^2$, they cannot be obtained as closed forms. But
for a limited class of functions of $H^2$, it was possible to
express the solutions in closed forms. In Ref.~\cite{kkl1}, the
closed form solution was obtained for the case of $1/H^2$. In Sec.
III of this paper, we obtained a closed form solution for $\log
(-H^2)$. It was also shown that the logarithmic Lagrangian allows
again the logarithmic Lagrangian in the dual picture. Without the
scalar-brane coupling, we anticipate that the duality symmetry may
play some role for stabilizing the logarithmic Lagrangian.

With the coupling of the scalar with the brane, it is possible to
restrict the form of solutions. For example, in the dual
description with the logarithmic function, the conditions become
Eqs.~(\ref{dsBC2}) and (\ref{dsBC1}). If the effective 4D
cosmological constant or the effective 4D curvature $\lambda$ is a
function of the vacuum expectation value of the scalar field
$\sigma$ at $y=0$, then the condition could determine $\lambda$.
In this case, one can hope to find a strong self-tuning solution.

\acknowledgments
This work is supported in part by the BK21 program
of Ministry of Education, the KOSEF Sundo Grant, and by the Office
of Research Affairs of Seoul National University.

\end{document}